\documentclass[prd,twocolumn,superscriptaddress,preprintnumbers,showpacs,nofootinbib,notitlepage,floatfix]{revtex4-1}
\usepackage{hyperref}
\usepackage{graphicx}
\usepackage{subfigure}
\usepackage{amsmath}
\usepackage{amssymb}
\usepackage{slashed}
\usepackage{amsfonts}
\usepackage{xcolor}
\usepackage{multirow}
\usepackage{array}
\usepackage{mwe}
\usepackage{mathrsfs}

\begin{document}

\title{Power corrections in the determination of heavy meson LCDAs: A renormalon-based estimation}

\author{Tu Guo} 
\affiliation{State Key Laboratory of Dark Matter Physics, School of Physics and Astronomy, Shanghai Jiao Tong University,  Shanghai 200240, China}
\author{Chao Han}
\email{chaohan@sjtu.edu.cn}
\affiliation{State Key Laboratory of Dark Matter Physics, School of Physics and Astronomy, Shanghai Jiao Tong University,  Shanghai 200240, China}
\author{Wei Wang}
\email{wei.wang@sjtu.edu.cn}
\affiliation{State Key Laboratory of Dark Matter Physics, School of Physics and Astronomy, Shanghai Jiao Tong University,  Shanghai 200240, China} 
\author{Jia-Lu Zhang}
\email{elpsycongr00@sjtu.edu.cn}
\affiliation{Tsung-Dao Lee Institute, Shanghai Jiao Tong University, Shanghai 201210, China}
\affiliation{State Key Laboratory of Dark Matter Physics, School of Physics and Astronomy, Shanghai Jiao Tong University,  Shanghai 200240, China}

\begin{abstract}
At leading power accuracy, the QCD light-cone distribution amplitudes (LCDAs) for a heavy meson can be matched onto the LCDAs in the framework of heavy-quark effective theory (HQET) through a factorization formula. We examine the power corrections to this factorization in the renormalon model, which can associate the power corrections originating from high-twist contributions to the divergent series in a matching kernel. Our analysis indicates that the dominant  power corrections originate from the virtual part of the vertex bubble chain diagrams, which generate poles at $w=n+\frac{1}{2}, \forall n\in \mathbb{N}$ and $w=1$ in the Borel plane. Employing phenomenological models for  both HQET and QCD LCDA,  we present a numerical  estimate. The results indicate that the power corrections in the peak region are approximately $22\%$ for the D meson and $7\%$ for the $\overline{\mathrm{B}}$ meson.  These findings showcase the magnitude and the potential importance of power corrections in achieving high-precision phenomenological predictions for heavy mesons.

\end{abstract}

\maketitle



\section{Introduction}
Heavy Quark Effective Theory (HQET) provides an essential framework for studying the dynamics of hadrons containing a heavy quark, enabling systematic analyses of the interplay between perturbative and non-perturbative QCD effects. One of the key objects in HQET is the light-cone distribution amplitude (LCDA) that encapsulates the non-perturbative structure of heavy-light mesons. The HQET LCDAs describe the momentum  carried by the light degrees of freedom of a heavy meson~\cite{Grozin:1996pq,Lange:2003ff,Braun:2003wx,Geyer:2005fb,Braun:2017liq,Kawamura:2001jm}. Unlike their counterparts in full QCD, HQET LCDAs focus solely on the soft physics associated with the light quark and gluons with the heavy quark mass $m_Q$ effectively decoupled. This separation greatly simplifies the description of hadronic states at a price of an $\Lambda_{\mathrm{QCD}}/m_Q$ power expansion. HQET LCDAs are essential to the precision of QCD phenomenology, especially in the study of exclusive 
B meson decays, where their use is grounded in the factorization theorem~\cite{Beneke:1999br,Beneke:2000ry,Lu:2000em}.
By providing a universal non-perturbative input for factorization theorems, the HQET LCDAs enable accurate predictions of decay rates and form factors~\cite{Keum:2000wi,Lu:2000em,Lu:2002ny,Khodjamirian:2005ea,Hwang:2010hw}, which are indispensable for examining the standard model and searching for new physics.

Many advancements have been made in the last few decades to explore the renormalization group evolution (RGE) and other perturbative properties of HQET LCDAs~\cite{Ball:1998ff,Ball:1998sk,Lee:2005gza,Kawamura:2010tj,Feldmann:2014ika,Braun:2015pha,Sun:2016avp,Binosi:2018rht,Braun:2019wyx,Serna:2020txe,Galda:2020epp,Hu:2024ebp,Wang:2024wwa}.  However, due to several difficulties, it is extremely difficult to obtain explicit distributions from first-principle calculations such as lattice QCD. First, HQET LCDAs contain lightcone Wilson lines, making direct lattice calculations impossible. Second, these LCDAs are defined with effective fields in HQET, and lattice QCD calculations of quantities involving HQET fields are very challenging~\cite{Mandula:1990fit}, often requiring sophisticated   schemes to address large discretization errors (see for instance Ref.~\cite{Horgan:2009ti}). More importantly, the co-existence of heavy quark fields and lightcone Wilson lines introduces a new type of divergence in HQET LCDAs, known as cusp divergence, which prevents the use of operator product expansion (OPE) and the method of determining distributions through moment calculations, as the non-negative moments become ill-defined~\cite{Braun:2003wx}.

Recently, a factorization theorem linking QCD LCDAs to the HQET LCDAs has been presented in Refs.~\cite{Zhao:2019elu,Ishaq:2019dst,Beneke:2023nmj}. The matching is performed within the context of effective field theory by integrating out the hard modes with virtualities of order $m_Q$.  These matching relations have been applied in the study of W decays to $\mathrm{B}$ mesons~\cite{Beneke:2023nmj,Ishaq:2024pvm}. Based on this relationship, a realistic method to extract  the HQET LCDA directly from lattice QCD has been proposed in Ref.~\cite{Han:2024min}, and first attempts on the lattice can be found in Ref.~\cite{Han:2024yun}.

Despite of this great success, it should be noticed that
in the procedure there are  power-suppressed corrections in terms of $\Lambda_{\mathrm{QCD}}/m_Q$, which need to be systematically accounted for to achieve high precision. These corrections are particularly important in the context of ongoing and forthcoming  experiments, where the high precision data demands corresponding theoretical accuracy.

One of the key challenges in accounting for power corrections in the matching procedure arises from their non-perturbative origin. In addition, accurate calculation of these corrections necessitates the inclusion of higher-twist LCDAs, which are currently inaccessible from  lattice QCD \cite{Zhang:2017bzy,Chen:2016utp}. 
To address this limitation, the renormalon model offers a valuable framework for estimating these corrections~\cite{Beneke:1994sw,Beneke:1996gx,Beneke:1998ui,Beneke:2000kc,Han:2024cht,Braun:2024snf}. Consequently, this approach has been successfully applied to various physical processes and quantities governed by factorization theorems~\cite{Beneke:1994rs,Beneke:1995pq,Scimemi:2016ffw,Braun:2018brg,Liu:2020rqi,Caola:2022vea,Schindler:2023cww,Mikhailov:2023gmo,Liu:2023onm,Han:2024cht,Braun:2024snf}.
Renormalons are associated with the divergence of the perturbative series in QCD. The IR and UV renormalons correspond to the infinitely large and small momentum limits in the loop integral, respectively. When combined with the framework of OPE, the renormalon model provides a systematic way to account for higher-power corrections. Within this renormalon model, corrections are estimated using the bubble chain diagrams. By incorporating this model, we provide a reasonable estimate of the impact of power corrections on the leading-twist factorization formula.

In this paper, we use the renormalon model to estimate the power corrections arising from matching QCD LCDAs to HQET LCDAs.
Specifically, we provide a perturbative calculation of the renormalon contribution, focusing on the evaluation of the bubble chain diagrams. In our analysis, we employ two methods: one involves performing a direct calculation of the bubble chain diagrams, while the other utilizes an effective Borel transformed gluon propagator as a direct replacement.
We find that the poles arise exclusively from the virtual part of vertex diagrams, with the gluon propagator replaced by a bubble chain. These poles are located at \( w = n + \frac{1}{2} \) for all \( n \in \mathbb{N} \), and at \( w = 1 \) in the Borel plane, corresponding to a pole configuration distinct from other HQET scenarios.

As a numerical estimate, 
we then employ phenomenological models for QCD and HQET LCDAs for the \( \mathrm{D} \) meson and \( \overline{\mathrm{B}} \) meson to calculate  the magnitude of the corrections predicted by the renormalon model. In this framework, we find that the corrections are approximately $22\%$ for the $\mathrm{D}$ meson and $7\%$ for the $\overline{\mathrm{B}}$ meson, which aligns with the naive  expectation based on the ratio $\Lambda_{\text{QCD}} / m_Q$. Through this explicit analysis, we aim to contribute to ongoing efforts in refining theoretical models in heavy quark  physics.

The remainder of this paper is organized as follows.
Section II provides a brief review of the theoretical foundations, including the boosted HQET,  HQET factorization, and the renormalon model.
Section III explores the application of the renormalon model for estimating power corrections in HQET factorization, with a detailed calculation procedure presented.
Section IV focuses on a numerical analysis of power corrections for both the \(\mathrm{D}\) meson and the $\overline{\mathrm{B}}$ meson.
Section V concludes the paper with a summary.
All results of the relevant bare bubble chain diagrams are provided as a complement in Appendix A.

\section{Theoretical Foundation}
\subsection{A Brief Introduction to Boosted HQET}

Boosted Heavy Quark Effective Theory (bHQET) extends the traditional HQET framework to accommodate scenarios involving heavy quarks with large velocities. While HQET simplifies the dynamics of heavy quarks at rest or moving slowly relative to a chosen reference frame, bHQET is specifically designed for high-energy processes where heavy quarks are relativistic, as frequently encountered in collider physics and jet dynamics \cite{Gracia:2021nut,Beneke:2023nmj,Clavero:2024yav}.
This leads to the necessity of considering both the heavy quark and the large momentum scale within this theory.

The power-counting in bHQET introduces two small parameters. The parameter, \( \lambda = \Lambda_{\text{QCD}} / m_Q \), originates from HQET and represents the suppression from $m_Q$. The second parameter, \( b = m_H / Q \), captures the effects of the large boost imparted to the heavy quark. The effective Lagrangian in bHQET is constructed by performing a field redefinition and matching the boosted quark dynamics to full QCD in the large-velocity limit. 
From a practical point of view, this framework can be regarded as a combination of HQET and Soft-Collinear Effective Theory (SCET)~\cite{Bauer:2000yr,Bauer:2001yt,Beneke:2002ph,Beneke:2002ni}. 

One can approach this either by constructing the theory based on HQET and subsequently incorporating SCET, or conversely, by starting with SCET and then adding the elements of HQET.
Specifically, the leading heavy field is defined as
\begin{align}
    h_{n}(x)= \sqrt{ \frac{2}{n_{+}v} } e^{ i m_{Q}v \cdot x } \frac{\not n_{-} \not n_{+}}{4 } Q(x).
\end{align}
This definition provides a systematic framework for power counting. Additionally, it reproduces the same results as HQET at leading order
\begin{align}&
    \mathcal{L}_{\text{HQET}}= \bar{h}_{v}(x)i v \cdot D h_{v}(x) = \notag
    \\&
    \bar{h}_{n}(x) i v \cdot D \frac{\not n_{+}}{2} h_{n}(x) (1+\mathcal{O}(\lambda))= \mathcal{L}_{\text{bHQET}}(1+\mathcal{O}(\lambda)).
\end{align}
This Lagrangian incorporates interactions with collinear and ultrasoft gluons, which are crucial for accurately describing high-energy processes.

In bHQET, the LCDA for a 
heavy pseudoscalar meson is defined as~\cite{Grozin:1996pq}
\begin{align}
\begin{aligned}
\varphi^+(\omega,\mu) = & \frac{1}{i\tilde{f}_H(\mu)m_H} \int_{-\infty}^{+\infty} \frac{d\eta}{2\pi} e^{i\omega n_+ \cdot v \eta} \\
& \times \left\langle 0 \left| \bar{q}(\eta n_+) \not n_+ \gamma_5 W_c(\eta n_+,0) h_n(0) \right| H(v) \right\rangle,
\end{aligned}
\end{align}
where \( n_\pm^\mu = (1, 0, 0, \mp 1)/\sqrt{2} \) are unit lightcone vectors, \( q(x) \) represents a light quark field and $\tilde{f}_H(\mu)$ is the decay constant for the heavy meson in HQET. 
The momentum of the meson is expressed as \( P_H^\mu = P_H^{+} n_-^\mu + \frac{m_H^2}{2 P_H^+} n_+^\mu \), while the heavy quark momentum is parametrized as \( P_Q^\mu = m_Q v^\mu + k^\mu \), where \( v \sim (1/b, 1, b) \) and \( k \sim (1/b, 1, b) \Lambda_{\text{QCD}} \) is the residual momentum.
To ensure gauge invariance, a Wilson line \( W_c(x, y) \) is introduced. Its explicit form is given by
\begin{align}
W_c(x, y) = \mathcal{P} \exp \left[ i g \int_0^1 \mathrm{~d} t \, (x-y)_\mu A^\mu(t x + (1-t) y) \right].
\end{align}

\subsection{Factorization of LCDAs into bHQET LCDAs}

Recently, the factorization of QCD LCDA to bHQET LCDA is established~\cite{Beneke:2023nmj}, which provides an ab initio method to calculate the bHQET LCDA.

The QCD LCDA is defined as
\begin{align}
\begin{aligned}\phi(u,\mu;m_H)&=\frac{1}{if_H}\int_{-\infty}^{+\infty}\frac{d\tau}{2\pi}e^{iyP_H\tau n_+}\\\times&\langle0\left|\bar{q}(\tau n_+)\not n_+\gamma_5 W_c(\tau n_+,0)Q(0)\right|H(P_H)\rangle ,\end{aligned}
\end{align}
where $f_H$ is the decay constant in QCD. $f_H$ and $\tilde{f}_H(\mu)$ are related through
\begin{align}
f_H=\tilde{f}_H(\mu)\left(1-\frac{\alpha_sC_F}{4\pi}\big(\frac{3}{2}\ln\frac{\mu^2}{m_Q^2}+2\big)+\mathcal{O}\left(\alpha_s^2\right)\right).
\end{align}
\subsubsection*{Power Counting in Peak and Tail Regions}

The QCD LCDA \( \phi(u,\mu;m_H) \) describes the light-cone momentum fraction \( u \) carried by the light anti-quark in the meson. Due to the large mass \( m_Q \) of the heavy quark, the LCDA exhibits distinct scalings in two regions:
\begin{itemize}
    \item \textbf{Peak Region (\( u \sim \lambda  \))}: Here, the LCDA scales as \( \phi(x) \sim \lambda^{-1} \).
    \item \textbf{Tail Region (\( u \sim 1 \))}: In this regime, the LCDA scales as \( \phi(x) \sim 1 \).
\end{itemize}
These scalings arise from the normalization condition and the typical momentum distribution of the light degree of freedom. To address these regions systematically, separate matching calculations are performed before merging them to describe the intermediate region.

\begin{figure}
    \centering  
\includegraphics[width=1\linewidth]{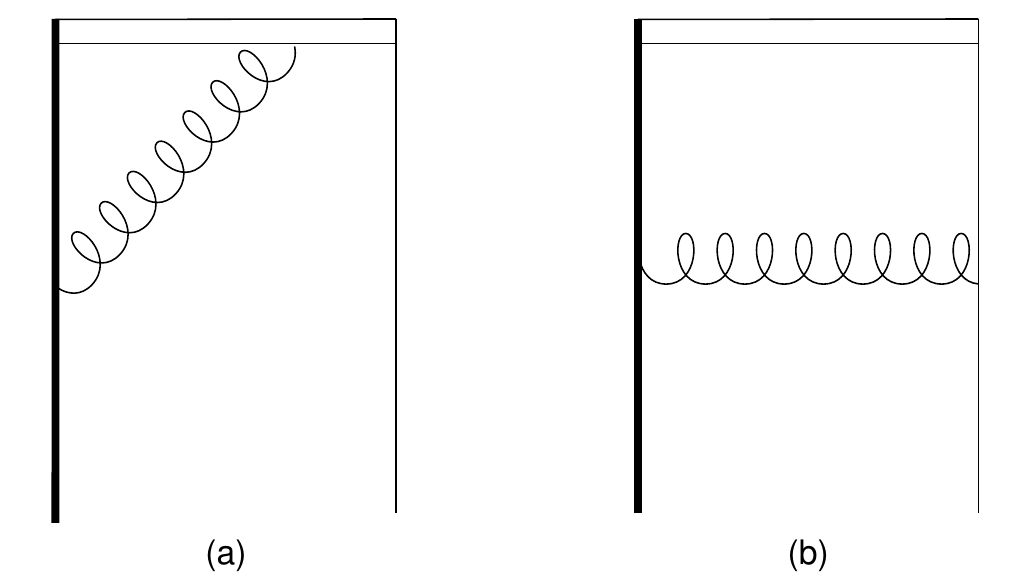}
\caption{The one-loop diagrams for the QCD and HQET LCDAs are shown. The thick line represents the heavy quark field, and the thin line represents the light anti-quark. }
\label{fig:diagram}
\end{figure}

\subsubsection*{Matching in the Peak Region}
In this paper, we will focus solely on the peak region, which serves as the relevant domain to describe the distribution at  $u\sim \lambda$.
In the peak region, the matching equation takes the form:
\begin{align}
\label{peak_matching}
\phi(u,\mu;m_H) = \frac{\tilde{f}_H(\mu)}{f_H} \int_0^\infty d\omega \, J_p(u, \omega) \varphi^+(\omega),
\end{align}
where the jet function \( J_p(u, \omega) \) is perturbatively calculable. 
The one-loop jet function \( J_p(u, \omega) \) has a simple structure due to cancellations between hard-collinear and soft-collinear regions:
\begin{align}
J_p(u, \omega) = \delta\left(u - \frac{\omega}{m_H}\right) \theta(m_H - \omega) \left(1 + \frac{\alpha_s C_F}{4\pi} J_{\text{peak}}^{(1)} \right)
\end{align}
where, in dimensional regularization and $\overline{\mathrm{MS}}$ (modified minimal subtraction) scheme, 
\begin{align}
J^{(1)}_{\mathrm{peak}}=1+\frac{\alpha_sC_F}{4\pi}\bigg(\frac{1}{2}\mathrm{ln}^2\frac{\mu^2}{m_H^2}+\frac{1}{2}\mathrm{ln}\frac{\mu^2}{m_H^2}+\frac{\pi^2}{12}+2\bigg).
\end{align}
This result simplifies Eq.~(\ref{peak_matching}) to
\begin{align}
\label{matching}
\phi\left(u,\mu;m_H\right)=&\frac{\tilde{f}_H(\mu)}{f_H}J_\text{peak}^{(1)}m_H\varphi^+\left(um_H,\mu\right)+\notag\\
&\frac{1}{m_Q^2}\phi_2(u,\mu;m_H)+...\, ,
\end{align}
where $\phi_2(u,\mu;m_H)$ stands for the subleading-twist operators in HQET that contributes the QCD LCDA. 

The one-loop diagrams that need to be calculated are shown in Fig.~\ref{fig:diagram}. It is worth noting that, in the peak region, only the virtual part of Fig.~\ref{fig:diagram}(a) contributes. This can be demonstrated either through the expansion-by-region strategy or via explicit calculation.

\subsection{Renormalon model}

As shown in Eq.~(\ref{matching}), higher-twist operators also contribute to the QCD LCDAs, which underestimates the extraction of HQET LCDAs by the power correction in the form $\displaystyle\frac{\Lambda_{\mathrm{QCD}}^{n}}{(m_Q)^{n}}$.  
Several approaches exist to account for higher-twist contributions.
One approach is to directly identify the matrix elements of higher-twist operators that contribute to the corrections, as was done for the HQET LCDA~\cite{Neubert:1992fk,Lee:2005gza,Braun:2017liq}.
Alternatively, the renormalon model can be used to estimate higher-twist contributions, an approach that has been widely applied in various works~\cite{Beneke:1995pq,Gehrmann:2012sc,Scimemi:2016ffw,Braun:2018brg,FerrarioRavasio:2020guj,Liu:2020rqi,Gracia:2021nut,Caola:2022vea,Makarov:2023ttq,Schindler:2023cww,Zhang:2023bxs,Schindler:2023cww,Mikhailov:2023gmo,Makarov:2023uet,Liu:2023onm,Han:2024cht,Braun:2024snf}.
In this paper, we adopt the renormalon model to estimate the higher-twist contributions to the heavy meson LCDAs.

The renormalon approach to power corrections is based on the observation that operators of different twists mix under renormalization.
If one can employ cutoff regularization in the perturbative calculation of QCD LCDAs and HQET LCDAs, the mixing is manifested. Since the matching kernel represents the physics in between the heavy quark mass $m_Q$ and the cutoff scale $\mu_F$, it can have power dependence on the $\mu_F$ and take the general form
\begin{align}
J_{\mathrm{peak}}=1+\sum_{n=0}^{\infty} c_n(\mu_F) a_s^{n+1}-\sum_{n=1}^{\infty}\left(\frac{\mu_F}{m_Q}\right)^nD_{\mathcal{Q}n},
\end{align}
where $a_s=\displaystyle \frac{g^2}{(4\pi)^{2}}$, and $c_n(\mu_F)$ depend logarithmically on  $\mu_F$. The terms $D_{\mathcal{Q}n}$ represent the power dependence on $\mu_F$ of the matching kernel. Since the left-hand side of Eq.~(\ref{matching}) is independent of the factorization scale, there must be a term \begin{align}
\tilde{\phi}_2=\mu_F^2\frac{\tilde{f}_H(\mu)}{f_H}D_{\mathcal{Q}_2}\varphi^+(um_H,\mu)
\end{align} 
in $\phi_2$ that shares the same power dependence on $\mu_F$ to cancel the contribution from $D_{\mathcal{Q}_2}$.  If the subleading-twist operators are dominated by their UV contributions---a feature referred to as ``ultraviolet dominance''~\cite{Beneke:1997sr,Beneke:1998ui}, and thus by $\tilde{\phi}_2$, the term $D_{\mathcal{Q}_2}$ will provide a reasonable approximation for the total contribution from subleading-twist operators.

However, in the context of dimensional regularization, which is more commonly used in practical calculations, the power dependence of the matching kernel on the factorization scale does not explicitly appear. Instead, the perturbative series \( \sum_{n=0}^{\infty} c_n(\mu_F) a_s^{n+1} \) diverges due to a factorial growth in $c_n(\mu_F)$. A detailed analysis in \cite{Beneke:1998ui} demonstrates that the factorial growth of the coefficients at large 
$n$ arises from contributions of extreme loop momenta, suggesting their relation to the power dependence on the factorization scale.  To assign a well-defined meaning to the divergent series, one can employ Borel summation techniques.

For the matching kernel, the Borel transform takes the following form:
\begin{align}
B[T](w) = \sum_{k=0}^{\infty} \frac{c_k}{k!} \left( \frac{w}{\beta_0} \right)^k.
\end{align}
The Borel integral
\begin{align}
\label{borel}
T(\alpha_s) = \frac{1}{\beta_0} \int_0^{\infty} d w \, e^{-w / \left( \beta_0 a_s \right)} B[T](w) 
\end{align}
then yields the asymptotic series \( \sum_{n=0}^{\infty} c_n a_s^{n+1} \) given above. Since the factorial growth of the series results in singularities in the Borel plane, the integral cannot be performed when these poles lie on the positive half-axis. To circumvent these singularities, one can deform the integration contour. The freedom in selecting the contour leads to the so-called renormalon ambiguity.

\begin{figure}
    \centering
\includegraphics[width=0.8\linewidth]{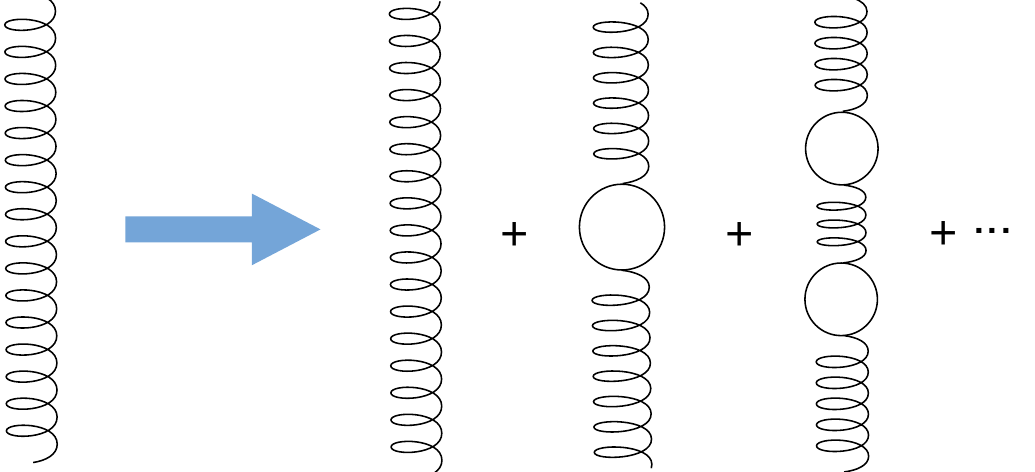}
\caption{The bubble chain diagrams correspond to changing a gluon propagator to gluon propagators dressed with fermion loops.}
\label{fig:bubble}
\end{figure}

The renormalon model is based on the expectation that the ambiguity is of the same order as the contribution from higher-twist operators, allowing it to serve as a parameterization of these higher-twist effects.  

In general, the full factorial growth in the matching kernels cannot be directly traced back to individual Feynman diagrams. However, it has been argued that the dominant contributions arise from a specific class of integrals, commonly referred to as ``bubble chain diagrams''~\cite{Gross:1974jv,Lautrup:1977hs,Zichichi:1977gu,Beneke:1998ui}. These are computed by inserting multiple fermion loops into a one-loop diagram, as shown in Fig.~\ref{fig:bubble}, and then applying the replacement
\begin{align}
    -\frac{2}{3} n_f \to \beta_0 = \frac{11}{3}N_c - \frac{2}{3}n_f. 
\end{align}
This  procedure is referred to as ``naive nonabelianization''~\cite{Broadhurst:1994se,Beneke:1994qe}.

To compute the bubble chain, we employ two methods~\cite{Beneke:1994sw}.  

\begin{itemize}
    \item \textbf{Two-Step Renormalization:}
    This approach begins by considering all possible configurations in which some fermion loops are replaced by their corresponding counterterms. After renormalizing all fermion loops, we proceed to handle the remaining UV divergence and transform the result into the Borel plane. This procedure will be described in detail in the following section.

    \item \textbf{Borel-Transformed Propagator:}  
    Alternatively, we directly use the Borel-transformed bubble chain propagator, given by  
    \begin{align}  
    \label{eff-pro}  
        B[a_s D^{AB}(k)](w) = i\delta^{AB} \left(e^{5/3} \mu^2 \right)^w \frac{g_{\mu\nu} - \frac{k_\mu k_\nu}{k^2}}{(-k^2 - i\epsilon)^{1+w}},  
    \end{align}  
    and then perform the remaining integral.  
\end{itemize}

Both methods provide complementary insights into the calculation and serve as cross-validations of the results.

\section{Power Corrections}

In our specific case, the bubble chain diagrams correspond to the one-loop diagrams depicted in Fig.~\ref{fig:diagram} 
(a), where the gluon propagator is modified to  include \( n \) fermion loops. All other diagrammatic contributions cancel out in both the QCD and bHQET cases. This outcome is consistent with expectations, as only the gluon propagator is modified compared to the one-loop diagrams. Thus, this modification does not impact the application of the expansion-by-region technique. Explicit results for the bubble chain diagrams, both for the QCD LCDA and HQET LCDA cases, are provided in Appendix~\ref{app1}. To calculate the matching kernel, we need the scale of the partonic state:
$$
\begin{aligned}
p^{\mu}_{Q} &= m_{Q} v^{\mu} = \bar{s} n_{+} p_{H} \frac{n^{\mu}_{-}}{2}+ \frac{m_{Q}^{2}}{\bar{s}n_{+}p_{H}} \frac{n^{\mu}_{+}}{2}, \\
p^{\mu}_{q} &= n_{+} p_{q} \frac{n^{\mu}_{-}}{2}=s n_{+}p_{H} \frac{n^{\mu}_{-}}{2},     
\end{aligned}
$$ where $p_Q^\mu(p_q^\mu)$ is the momentum of heavy (light) quark,  $n_{+}p_{H} = m_{Q}v^{\mu}=\bar{s}n_{+} p_{H} \frac{n_{-}^{\mu}}{2}+\frac{m^{2}_{Q}}{\bar{s}n_{+}p_{H}} \frac{n_{+}^{\mu}}{2}$ and $\bar{s}\equiv 1-s$.

In this section, we calculate Fig.~\ref{fig:diagram} 
(a) in the coordinate space and then perform the Fourier transformation to the momentum space. The calculation requires several useful formulas that we now collect. 

The massive quark propagator can be written as
\begin{align}
G(x-y)=&\frac{(-i)^{d/2}}{2\pi^{d/2}}\int_0^\infty d\alpha \alpha^{d/2-1}e^{-i\alpha (x-y)^2-i\frac{m^2}{4\alpha}}\notag\\
&\times (i(x\!\!/-y\!\!/)+\frac{im}{2\alpha}). 
\end{align} 
The gluon propagator with $n$ fermion loops can be written as 
\begin{align}
    \Delta_n^{\mu\nu}(x-y)=\frac{-1}{4\pi^{d/2}}\frac{\Gamma(1-(n+1)\epsilon)}{\Gamma(1+n\epsilon)}\frac{g_{\mu\nu} R_\epsilon^n}{(-(x-y)^2+i\epsilon)^{1-(n+1)\epsilon}},
\end{align}
where the color factor is omitted, and
\begin{align}
R_\epsilon = -\frac{2n_f}{3}\frac{g^2}{(4\pi)^{d/2} \epsilon}\frac{6\Gamma(1+\epsilon)\Gamma^2(2-\epsilon)}{4^\epsilon \Gamma(4-2\epsilon)}.
\end{align} 
During the calculation, we only need the Schwinger parametrized form of it
\begin{align}
    \Delta_n^{\mu\nu}(x-y)=&-\frac{(-i)^{1-(n+1)\epsilon}}{4\pi^{d/2}}\frac{g_{\mu\nu}R^n_\epsilon}{\Gamma(1+n\epsilon)}\notag\\
    &
    \times \int_0^\infty d\alpha \alpha^{-(n+1)\epsilon}e^{-i \alpha(x-y)^2},
\end{align}
where we used the Schwinger parametrization of propagators
\begin{align}
\frac{1}{(-z^2+i\epsilon)^{N+1}}=\frac{(-i)^{N+1}}{\Gamma(N+1)}\int^\infty_0d\sigma \sigma^Ne^{-i\sigma z^2}.
\end{align} 

To Fourier transform the HQET LCDA in coordinate space to momentum space, we need the Fourier transformation of the following type: 
\begin{align}
&\int \frac{dt}{2\pi}e^{i x t }(it+\epsilon)^{2(n+1)\epsilon-1}\notag
\\&=\frac{1}{\Gamma(1-2(n+1)\epsilon)}\int\frac{dt}{2\pi}\int_0^\infty d\sigma \sigma^{-2(n+1)\epsilon}e^{-i \sigma t}e^{ixt}\notag\\
&=\frac{1}{\Gamma(1-2(n+1)\epsilon)}\int_0^\infty d\sigma \sigma^{-2(n+1)\epsilon}\delta(\sigma-x)\notag\\
&=\frac{1}{\Gamma(1-2(n+1)\epsilon)}\frac{1}{x^{2(n+1)\epsilon}}\theta(x).
\end{align}

\subsubsection{Bare bubble chain calculation}

Now, we explicitly calculate Fig.~\ref{fig:diagram} 
(a) in coordinate space. For convenience, we use the substitution  $z\equiv n_+ t$. The integral expression for Fig.~\ref{fig:diagram} 
(a)
in coordinate space is
\begin{widetext}
\begin{align}
\phi_a(z)=\frac{(-i)^{-(2+n) \epsilon } g^2 R_{\epsilon }^n}{8
   \pi ^{4-2 \epsilon } \Gamma (1+n \epsilon )}&\int d^dk\int_0^1 dt' \int_0^\infty d\sigma_1\int_0^\infty d\sigma_2 \sigma _2^{1-\epsilon } \sigma _1^{-((n+1) \epsilon
   )}\int d^d\eta  \notag\\
   &
   \times \exp \left(- i \left( \eta ^2 \sigma
   _2+\frac{m^2}{4\sigma _2}+ \sigma _1 (\eta
   -t' z)^2\right)-i k\eta\right)
   \bar{q}(z)n\!\!\!/_+(\eta\!\!\!/+\frac{m}{2\sigma_2})z\!\!/ \gamma_5q(k)  \notag\\
   =\frac{(-i)^{-(2+n) \epsilon } g^2 R_{\epsilon }^n}{8
   \pi ^{4-2 \epsilon } \Gamma (1+n \epsilon )}&\int d^dk\int_0^1 dt' \int_0^\infty d\sigma_1 \int_0^\infty d\sigma_2 \sigma _2^{1-\epsilon } \sigma _1^{-((n+1) \epsilon
   )}\int d^d\eta \notag\\
   &
   \times\exp \left(- i \left( \eta ^2 \sigma
   _2+\frac{m^2}{4\sigma _2}+ \sigma _1 (\eta
   -t' z)^2\right)-i k\eta\right)
   \bar{q}(z)n\!\!\!/_+\eta\!\!\!/z\!\!/ \gamma_5q(k).
\end{align}	
The second equality arises from $n\!\!\!/_+z\!\!\!/=tn\!\!\!/_+n\!\!\!/_+=tn_+^2=0$.
We complete the square for the exponential:
\begin{align}
    - i \left( \eta ^2 \sigma
   _2+\frac{m^2}{4\sigma _2}+ \sigma _1 (\eta
   -t' z)^2\right)-i k\eta\to-\frac{1}{4} i \left(-\frac{\left(k-2 \sigma _1 t'
   z\right){}^2}{\sigma _1+\sigma
   _2}+\frac{m^2}{\sigma _2}+4 \sigma _1 t'^2
   z^2\right)-i \left(\sigma _1+\sigma _2\right)
   \left(\eta +\frac{k-2 \sigma _1 t' z}{2
   \left(\sigma _1+\sigma _2\right)}\right){}^2.
\end{align} 
We introduce a change of variable $\xi=\eta +\frac{k-2 \sigma _1 t' z}{2
   \left(\sigma _1+\sigma _2\right)}$, which transforms the integral to
\begin{align}
\phi_a(z)=&\frac{(-i)^{-(2+n) \epsilon } g^2 R_{\epsilon }^n}{8
   \pi ^{4-2 \epsilon } \Gamma (1+n \epsilon )}\int d^dk\int_0^1 dt' \int_0^\infty d\sigma_1\int_0^\infty d\sigma_2 \sigma _2^{1-\epsilon } \sigma _1^{-((n+1) \epsilon
   )}\int d^d\xi  \notag\\
   &\times \exp \left(-\frac{1}{4} i \left(-\frac{\left(k-2 \sigma _1 t'
   z\right){}^2}{\sigma _1+\sigma
   _2}+\frac{m^2}{\sigma _2}+4 \sigma _1 t'^2
   z^2\right)-i \left(\sigma _1+\sigma _2\right)
   \xi^2\right)
   \bar{q}(z)n\!\!\!/_+(\xi\!\!/-\frac{k\!\!/-2 \sigma _1 t' z\!\!/}{2
   \left(\sigma _1+\sigma _2\right)})z\!\!/ \gamma_5q(k) \notag\\
    =&-\frac{(-i)^{-(2+n) \epsilon } g^2 R_{\epsilon }^n}{8
   \pi ^{4-2 \epsilon } \Gamma (1+n \epsilon )}\int d^dk\int_0^1 dt' \int_0^\infty d\sigma_1\int_0^\infty d\sigma_2 \sigma _2^{1-\epsilon } \sigma _1^{-((n+1) \epsilon
   )}\int d^d\xi  \notag\\
   &\times \exp \left(-\frac{1}{4} i \left(-\frac{\left(k-2 \sigma _1 t'
   z\right){}^2}{\sigma _1+\sigma
   _2}+\frac{m^2}{\sigma _2}+4 \sigma _1 t'^2
   z^2\right)-i \left(\sigma _1+\sigma _2\right)
   \xi^2\right)
   \bar{q}(z)n\!\!\!/_+\frac{k\!\!/}{2
   \left(\sigma _1+\sigma _2\right)}z\!\!/ \gamma_5q(k).
\end{align}	    
The second equality follows from the fact that the integrand vanishes if it is an odd function in \( \xi \) and \( z\!\!\!/z\!\!\!/ = 0 \).

After integrating the variable $\xi$:
    \begin{align}
\phi_a(z)=&-\frac{(-i)^{-1-(n+1) \epsilon } g^2 R_{\epsilon }^n}{8
   \pi ^{2- \epsilon } \Gamma (1+n \epsilon )}\int d^dk\int_0^1 dt' \int_0^\infty d\sigma_1 \int_0^\infty d\sigma_2 \sigma _2^{1-\epsilon } \sigma _1^{-((n+1) \epsilon
   )}  \notag\\
   &\times \exp \left(-\frac{1}{4} i \left(-\frac{\left(k-2 \sigma _1 t'
   z\right){}^2}{\sigma _1+\sigma
   _2}+\frac{m^2}{\sigma _2}\right)\right)
   \bar{q}(z)n\!\!\!/_+\frac{k\cdot z}{
   \left(\sigma _1+\sigma _2\right)^{1+d/2}}\gamma_5q(k),
\end{align}	
we change the variables
$\{\sigma_1,\sigma_2\}\to\{\sigma=\frac{\sigma_1\sigma_2}{\sigma_1+\sigma_2},\beta_1=\frac{\sigma_2}{\sigma_1+\sigma_2}\}$ and integrate out $\sigma$ and $t'$:
\begin{align}
\phi_a(z)=&-\frac{(-i)^{-1-(n+1) \epsilon } g^2 R_{\epsilon }^n}{8
   \pi ^{2- \epsilon } \Gamma (1+n \epsilon )}\int d^dk\int_0^1 dt' \int_0^\infty d\sigma \int_0^1 d\beta_1 \beta _1^{n \epsilon +1} \sigma ^{-((n+1) \epsilon
   )-1}
  \notag \\
   &\times \exp \left(\frac{i \left(\beta _1-1\right) \left(4 k \sigma  t'
   z-\beta _1 m^2+m^2\right)}{4 \sigma }\right)
   k\cdot z \bar{q}(z)n\!\!\!/_+\gamma_5q(k)  \notag\\
=&-4^{(n+1) \epsilon } \frac{g^2 R_{\epsilon }^nm^{-2(n+1)\epsilon}\Gamma (n \epsilon +\epsilon )}{8
   \pi ^{2- \epsilon } \Gamma (1+n \epsilon )}\int d^dk\int d\beta_1 \beta _1^{n \epsilon +1}
   \notag\\&\times 
   (1-\beta_1)^{-1-2(n+1)\epsilon} \bar{q}(z)n\!\!\!/_+\gamma_5q(k)  \left(1-e^{i \left(\beta _1-1\right) k
   z}\right).
\end{align}	
\end{widetext}

In the following steps, we contract the quark fields with the partonic state and discard the tree-level matrix element
\begin{align}
\phi_a(z)=&-4^{(n+1) \epsilon } \frac{g^2 R_{\epsilon }^nm^{-2(n+1)\epsilon}\Gamma (n \epsilon +\epsilon )}{8
   \pi ^{2- \epsilon } \Gamma (1+n \epsilon )}\int d\beta_1 \beta _1^{n \epsilon +1}
   \notag\\&
      \times (1-\beta_1)^{-1-2(n+1)\epsilon}   e^{i (u-s) P_H^+ t}  \left(1-e^{i \left(\beta _1-1\right) \bar{s}
   P_H^+ t}\right).
\end{align}	
We only care about the virtual part, which corresponds to 
\begin{align}
\phi_a(z)=&-4^{(n+1) \epsilon } \frac{g^2 R_{\epsilon }^nm_Q^{-2(n+1)\epsilon}\Gamma (n \epsilon +\epsilon )}{8\pi ^{2- \epsilon } \Gamma (1+n \epsilon )}\int d\beta_1 \beta _1^{n \epsilon +1}
   \notag\\&
\times (1-\beta_1)^{-1-2(n+1)\epsilon}   e^{i (u-s) P_H^+ t},
\end{align}
and finally we arrive at the bare result of $n$ bubble diagram for the virtual part of Fig.~\ref{fig:diagram} 
(a) in momentum space
\begin{align}
\phi_a(u)&=\frac{1}{if_H}\int_0^\infty\frac{dt}{2\pi}e^{i u P_H^+t}\phi_{a}(n_+t)  \notag\\
&=i4^{(n+1) \epsilon } \frac{g^2 R_{\epsilon }^nm_Q^{-2(n+1)\epsilon}\Gamma (n \epsilon +\epsilon )}{8
   \pi ^{2- \epsilon }f_HP_H^+ \Gamma (1+n \epsilon )}
   \notag\\&
   \times \frac{\Gamma (-2 (n+1) \epsilon ) \Gamma (n \epsilon
   +2)}{\Gamma (2-(n+2) \epsilon )} \delta(u-s).
\end{align}

The bare diagram does not include a factor of $n!$, which leads to the factorial growth in the matching kernel. This factorial behaviour arises from the renormalization of the fermion loops.
\subsubsection{Renormalization of bubble chain diagram}

In the renormalization process, a bare diagram with 
$n$ fermion loops is replaced by a sum of diagrams where some of the fermion loops are substituted with their corresponding counter-terms. Then we discard the terms with $\epsilon^{-n}( n>0)$. This substitution introduces the factorial growth, as each permutation of fermion loops and counter-terms contributes to the $n!$ factor
\begin{widetext}
\begin{align}
f_H\phi_{a.ren}&=\sum_{k=0}^n \binom{n}{k}(\frac{-\beta_0}{\epsilon})^k\frac{i(4\pi)^\epsilon a_s^{k+1} 2^{-2 k \epsilon +2 n \epsilon +1} (n
   \epsilon -k \epsilon +1) m_Q^{2 \epsilon 
   (k-n-1)} R_{\epsilon }^{n-k} \Gamma (2 (k-n-1)
   \epsilon ) \Gamma ((-k+n+1) \epsilon )}{P_H^+ \Gamma
   ((k-n-2) \epsilon +2)} \delta(u-s)  \notag\\
   &=a_s^{n+1}(\frac{\beta_0}{\epsilon})^n\sum_{k=0}^n\binom{n}{k}(-1)^k\frac{ i2 (s-\epsilon +1) m_Q^{-2s}  \Gamma (-2s) \Gamma (s)(4\pi)^{s}}{P_H^+ \Gamma(-s-\epsilon+2)}f^{s/\epsilon-1}\delta(u-s), 
\end{align}
\end{widetext}
where $f=\frac{6 \Gamma (2-\epsilon )^2 \Gamma
   (\epsilon +1)}{\Gamma (4-2 \epsilon )}$ and $s=(n-k+1)\epsilon$.
To simplify the expression, one can set
\begin{align}
  F(\epsilon,s)&=\frac{ i2 (s-\epsilon +1) m_Q^{-2s}  \Gamma (-2s) \Gamma (s)(4\pi)^{s}}{P_H^+ \Gamma
   (-s-\epsilon+2)}f^{s/\epsilon-1},  \\
   F(\epsilon,s)&=\sum_{j=-2}^\infty \sum_{n=0}^\infty f^{[j]}_ns^j\epsilon^n, 
\end{align}
and use the formulas related to the second Stirling  numbers:
\begin{align}
 & 
\begin{aligned}
\sum_{k=0}^n\frac{(-1)^k\binom{n}{k}}{(n-k+1)^2}=\frac{(-1)^n}{n+1}H_{n+1},
\end{aligned} \\
 & 
\begin{aligned}
\sum_{k=0}^n\frac{(-1)^k\binom{n}{k}}{(n-k+1)}=\frac{(-1)^n}{n+1},
\end{aligned} \\
 & \sum_{k=0}^n(-1)^k\binom{n}{k}(n-k+x)^j=0,0\leq j\leq n-1,\forall x, \\
 & \sum_{k=0}^n(-1)^k\binom{n}{k}(n-k+x)^n=n!,\forall x. & 
\end{align}

Finally, one arrives at
\begin{align}
\label{psiren}
&
\phi_{ren}/a_s^{n+1}= \notag
\\&
    \beta_0^n
  \bigg(f_{n+2}^{[-2]}\frac{(-1)^n}{n+1}H_{n+1}+f_{i+1}^{[-1]}\frac{(-1)^n}{n+1}+f_0^{[n]}n!\bigg)\delta(u-s).
\end{align}  
One can observe that the last term in Eq.~(\ref{psiren}) contains a factorial growth in $n$.

\subsubsection{Renormalon Ambiguity}
The renormalon ambiguity corresponds to the residue of the Borel transform of matching kernel. The Borel transform of $\phi_{a,ren}$ can be written as
\begin{align}
    \mathcal{B}[f_H\phi_{a.ren}](w)&= R(w)+ \sum_{n=0}^{\infty} f_0^{[n]} w^n\notag\\
    &=\big(R(w)+ F(0, w)\big)\delta(u-s)
 \end{align}
 where $R(w)$ is an entire function in $w$, which corresponds to the first two terms in Eq.~(\ref{psiren}) and do not contribute to renormalon ambiguity.

 Putting back the color factor and $\mu_0^{2w}=\mu^{2w}(\frac{e^{\gamma_E}}{4\pi})^w$, we get the Borel transform of matching kernel
 \begin{align}
\mathcal{B}[&J_{\text{peak}}](w) = \notag\\&-i C_F R(w)\mu_0^{2w} -\frac{2C_F (w+1) \mu^{2w} \Gamma(-2w) \Gamma(w)}{m_Q^{2w} \Gamma(2-w) } e^{5w/3}.
\end{align}

\begin{@twocolumnfalse}
    \begin{center}
    \begin{figure*}
\label{pic_model}
\centering
\begin{minipage}[t]{0.47\textwidth}
\centering
\includegraphics[width=8cm]{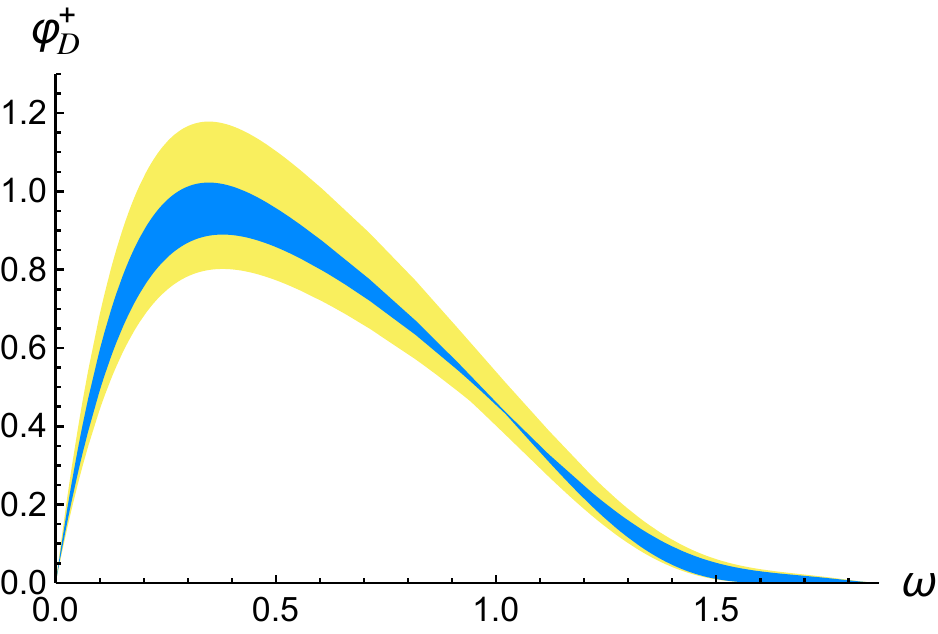}
\centerline{(a)}
\end{minipage}
\begin{minipage}[t]{0.47\textwidth}
\centering
\includegraphics[width=8cm]{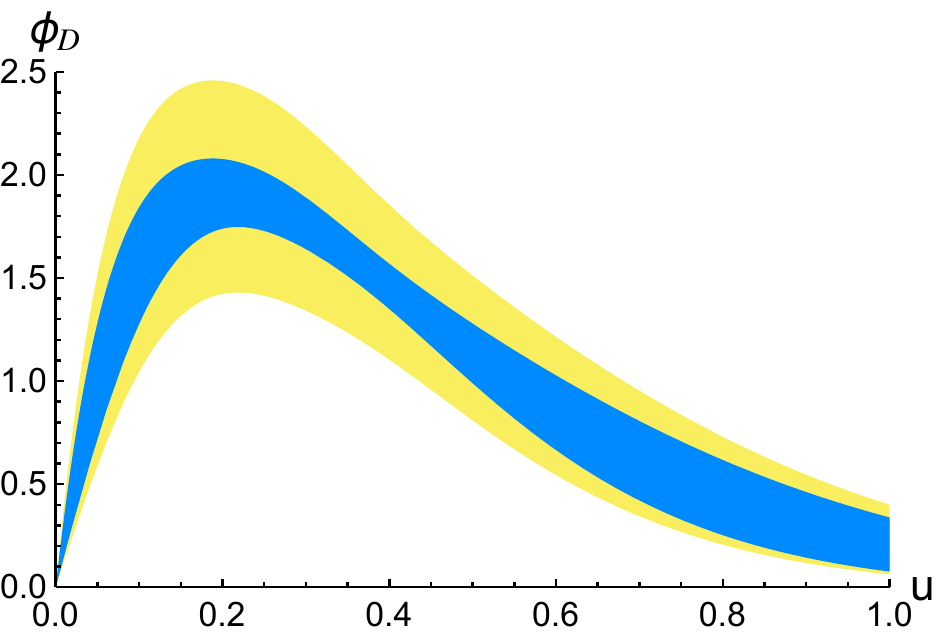}
\centerline{(b)}
\end{minipage}
\caption{(a) The HQET LCDA of the $\mathrm{D}$ meson with one-loop matching is shown. The blue error band represents the variation in the factorization scale, ranging from \( 1.6 \,\mathrm{GeV} \) to \( 3 \,\mathrm{GeV} \), while the yellow error band accounts for the renormalon ambiguity estimation. (b) The QCD LCDA of the $\mathrm{D}$ meson with one-loop matching is shown. The blue error band reflects the uncertainty associated with the HQET LCDA model, and the yellow error band represents the estimation of renormalon ambiguity.
 }
\label{D-meson}
\end{figure*}
    \end{center}
\end{@twocolumnfalse}

\begin{figure}
    \centering  
\includegraphics[width=1\linewidth]{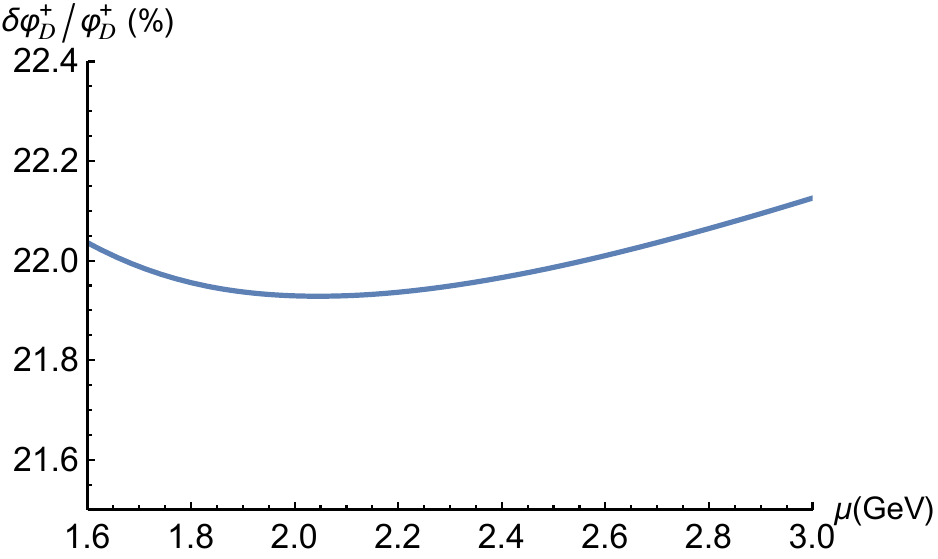}
\caption{The ratio of renormalon ambiguity to the one-loop matching result for the  $\mathrm{D}$  meson at different energy scales.}
\label{D-ratio}
\end{figure}
We then consider the Borel integral, as introduced in Eq.~(\ref{borel}). To estimate the size of the correction, it is sufficient to focus on the associated renormalon ambiguity
\begin{align}
&\delta \displaystyle J_{\text{peak}} \notag\\ = &\frac{1}{\beta_0}\mathrm{Res}\left|e^{-w / \left( \beta_0 a_s \right)}\frac{2C_F (w+1) \Lambda_{QCD}^{2w} \Gamma(-2w) \Gamma(w)}{m_Q^{2w} \Gamma(2-w) } e^{5w/3}\right| \notag\\
=  &\frac{1}{\beta_0}\mathrm{Res}\left|\frac{2C_F (w+1) \Lambda_{QCD}^{2w} \Gamma(-2w) \Gamma(w)}{m_Q^{2w} \Gamma(2-w) } e^{5w/3}\right|,
\end{align}
where the leading-order expression of $\alpha_s$ is used. The residue is evaluated at each point on the positive half-axis that exhibits a singularity. One can verify that there are singularities at $w=n+\frac{1}{2},\forall n\in \mathbb{N}$ and $w=1$.

\begin{figure}
    \centering  
\includegraphics[width=1\linewidth]{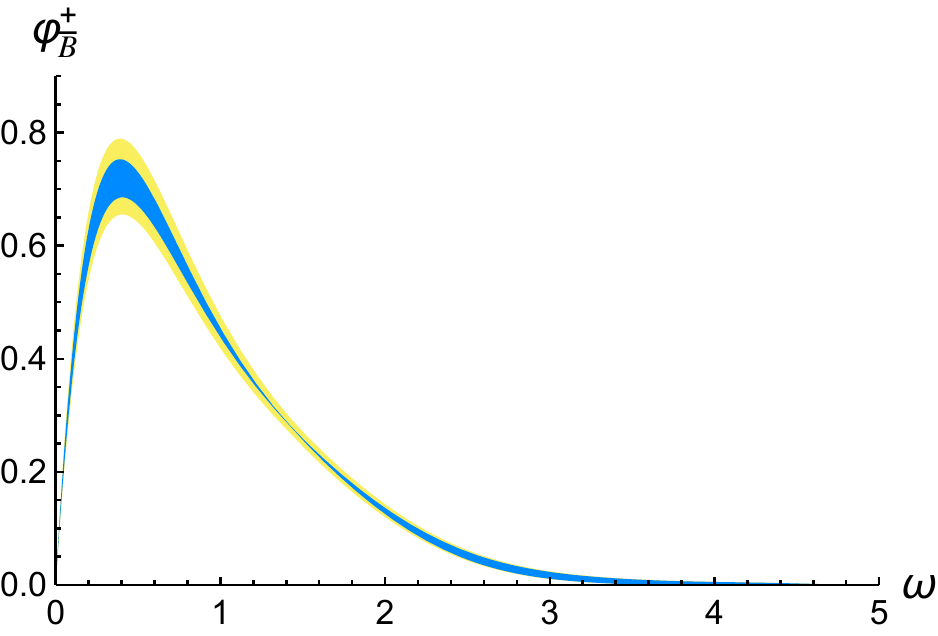}
\caption{The HQET LCDA of the $\overline{\mathrm{B}}$ meson with one-loop matching. The blue error band originates from the variation of the factorization scale from \( 4 \,\mathrm{GeV} \) to \( 6 \,\mathrm{GeV} \) and the yellow error band reflects the estimation of renormalon ambiguity.}
\label{B-meson}
\end{figure}


\section{Numerical Analysis}
Following the procedure outlined in the previous section, we analyze the power corrections up to second order in $\Lambda_{\mathrm{QCD}}/m_Q$ for the LCDAs of both $\mathrm{D}$ mesons and $\overline{\mathrm{B}}$ mesons. We employ established LCDA models to examine how power corrections influence the transition to HQET LCDAs. Additionally, we adopt phenomenological models for the HQET LCDA of the \( \mathrm{D} \) meson to estimate the power corrections when matching to the QCD LCDA. It should be noted that these analyses of power corrections may exhibit model dependence and can be improved once reliable results from lattice simulations of QCD LCDAs become available in the future.

\subsection{Power corrections to D meson LCDAs}
We parametrize the QCD LCDA using Gegenbauer polynomials:
\begin{align}
\phi(u,\mu) = 6u(1-u) \biggl[ 1 + \sum_{n=1}^{\infty} a_n(\mu) C^{(3/2)}_n(2u-1) \biggr]\,.
\end{align}

The Gegenbauer moments \( a_n(\mu) \) are determined through the integral
\begin{align}
a_n(\mu) &= \frac{2(2n+3)}{3(n+1)(n+2)} \int_0^1 du \, C_n^{(3/2)}(2u-1) \phi(u,\mu), 
\end{align}
and their evolution is governed by the relation
\begin{align}
\frac{a_n(\mu_h)}{a_n(\mu)} &= \left( \frac{\alpha_s(\mu_h)}{\alpha_s(\mu)} \right)^{\frac{\gamma_n}{2\beta_0}},
\end{align}
where the anomalous dimension \( \gamma_n \) is given by
\begin{align}
\gamma_n &= 2 C_F \left( 4 \sum_{k=1}^{n+1} \frac{1}{k} - \frac{2}{(n+1)(n+2)} - 3 \right).
\end{align}

For our numerical study, we adopt the meson mass \( m_D = 1.870 \,\mathrm{GeV} \). The numerical values of Gegenbauer moments are taken from Ref.~\cite{Beneke:2023nmj} and used as a phenomenological input. The Gegenbauer moments for the \( \mathrm{D} \) meson at \( \mu = 1.6 \,\mathrm{GeV} \) are
\begin{align}
a_n^D(1.6 \,\mathrm{GeV}) = \{-0.659, 0.206, -0.057, 0.036,&\notag\\
-0.004, -0.007, \dots\}&\,.
\end{align}

For the HQET LCDA of the \( \mathrm{D} \) meson, we adopt the model in \cite{LatticeParton:2024zko}:
\begin{align}
\varphi^+(\omega, \mu) &= \sum_{n=1}^{N} c_n \frac{\omega^n}{\omega_0^{n+1}} e^{-\omega/\omega_0} \notag \\
&= \frac{c_1 \omega}{\omega_0^2} \left[ 1 + c'_2 \frac{\omega}{\omega_0} + c'_3 \left( \frac{\omega}{\omega_0} \right)^2 + \cdots \right] e^{-\omega/\omega_0},
\end{align}
where \( \omega_0 = 0.32(15) \), \( c_1 = 0.63(44) \), \( c'_2 = 0.12(37) \), and \( c'_3 = 0.04(19) \).

In the left panel of Fig.~\ref{D-meson}, we illustrate the HQET LCDA for the \( \mathrm{D} \) meson, derived from the QCD LCDA model using one-loop matching. The blue error band reflects the variation in the factorization scale, which ranges from \( 1.6 \,\mathrm{GeV} \) to \( 3 \,\mathrm{GeV} \). Similarly, the right panel of Fig.~\ref{D-meson} displays the QCD LCDA for the \( \mathrm{D} \) meson, obtained from the HQET LCDA model through one-loop matching. In this case, the factorization scale is fixed at \( \mu = m_D \). The error band for the one-loop matching arises from the uncertainty in the HQET LCDA model. When the renormalon ambiguity is considered, the uncertainty band expands symmetrically, indicating a significant theoretical uncertainty in the one-loop matching.

In Fig.~\ref{D-ratio}, we further present the ratio of the power correction to the one-loop matching, \( \delta \varphi^+_D/\varphi^+_D \). Since the matching kernel does not rely on momentum fraction $u$, this ratio is independent of the momentum of the light degree of freedom. It is evaluated as a function of \( \mu \) to estimate the impact of power corrections within the renormalon model. In the peak region, where matching is performed at \( \mu \sim m_Q \), we find a correction of approximately \( 22\% \) for the \( \mathrm{D} \) meson. This aligns with our expectation that power corrections scale as \( \Lambda_{\text{QCD}}/m_c \), indicating a substantial effect.

\subsection{Power corrections to $\overline{\mathrm{B}}$ meson LCDAs}

As a complementary analysis, we estimate the power correction to the \( \overline{\mathrm{B}} \) meson HQET LCDA. The model-dependent results for Gegenbauer moments of the \( \overline{\mathrm{B}} \) meson QCD LCDA at \( \mu = 4.8 \,\mathrm{GeV} \) are given by~\cite{Beneke:2023nmj}:
\begin{align}
a_n^{\overline{\mathrm{B}}}(4.8 \,\mathrm{GeV}) = \{-1.082, 0.826, -0.513, 0.288, -0.157,& \notag\\
0.078, -0.030, 0.008, \dots\}&\,. \label{eq:B_LCDA_moments_Beneke:2023nmj}
\end{align}
It is crucial to emphasize that the Gegenbauer expansion for  \(\overline{\mathrm{B}} \) meson LCDAs suffers from slow convergence, primarily due to the distribution peak approaching the boundary. This limitation is explicitly manifested in the results of Eq.~\eqref{eq:B_LCDA_moments_Beneke:2023nmj}, where eight Gegenbauer moments are employed to model the amplitude. As shown in Ref.~\cite{Beneke:2023nmj}, these moments can yield highly asymmetric distributions for \( \overline{\mathrm{B}} \) meson LCDAs, a feature that can also be seen in the profiles depicted in Fig.~\ref{B-meson}.  However, since our analysis focuses on the estimate of power corrections,  the qualitative behavior of our results might not be significantly changed.

\begin{figure}
    \centering  
\includegraphics[width=1\linewidth]{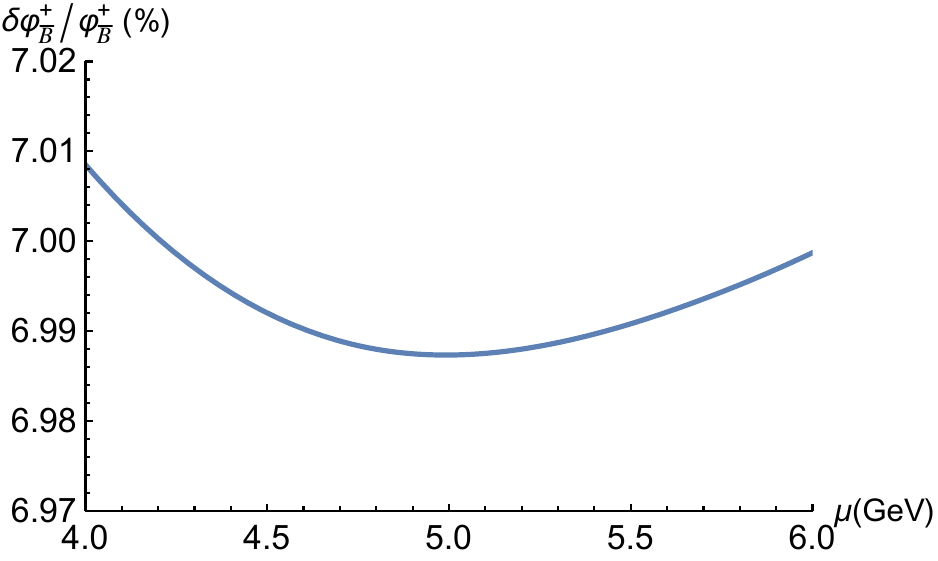}
\caption{The ratio of renormalon ambiguity to the one-loop matching result for the  \(\overline{\mathrm{B}}\) meson at different energy scales.}
\label{B-ratio}
\end{figure}

Fig.~\ref{B-meson} shows the HQET LCDA for the \( \overline{\mathrm{B}} \) meson, incorporating both one-loop matching and renormalon ambiguity as power corrections. The factorization scale is varied between \( 4 \,\mathrm{GeV} \) and \( 6 \,\mathrm{GeV} \). The ratio of the power correction to the one-loop matching is depicted in Fig.~\ref{B-ratio}, where a correction of about \( 7\% \) for the \( \overline{\mathrm{B}} \) meson is observed. 

\begin{figure}
    \centering  
\includegraphics[width=1\linewidth]{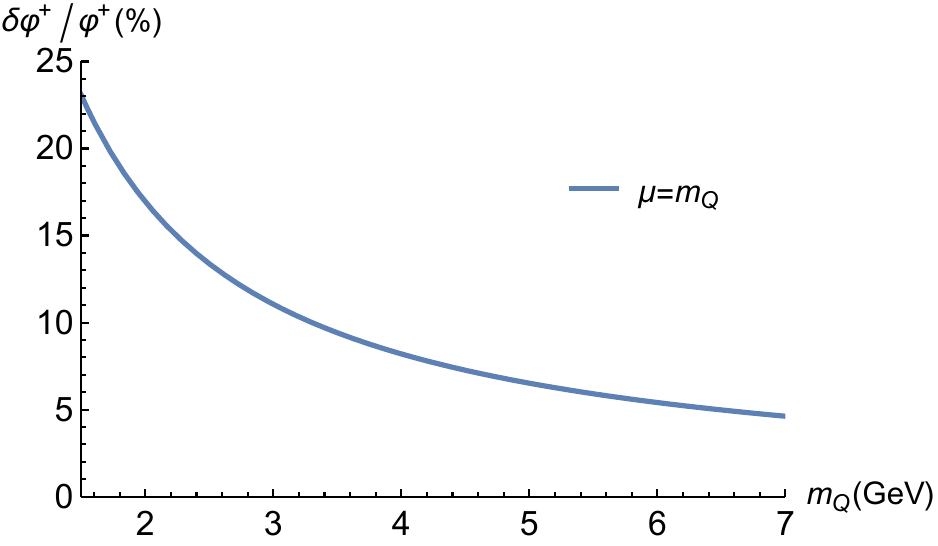}
\caption{The ratio of renormalon ambiguity to one-loop matching with the factorization scale set at \( \mu = m_Q \).}
\label{mass_evo}
\end{figure}

In Fig.~\ref{mass_evo}, we present the mass dependence of the ratio of power correction to the one-loop matching result, evaluated at \( \mu = m_Q \). As the meson mass decreases, the correction becomes increasingly significant, illustrating the limitations of the renormalon model. This highlights the necessity of explicit calculations for higher-twist contributions to achieve a more precise understanding.

Furthermore, these results can be extended to vector mesons using heavy-quark spin symmetry~\cite{Deng:2024dkd}. Although the matching relations for LCDAs with different spin structures differ at $\mathcal{O}(\Lambda_{\mathrm{QCD}}/m_Q)$, the estimation of higher-twist contributions within the renormalon model remains unchanged.
This further emphasizes the need for direct computations of higher-twist effects in order to improve precision.

\section{Summary}
In this paper, we estimate the power corrections involved in the matching between the heavy meson QCD LCDA and the HQET LCDA using the renormalon model.
Specifically, a perturbative calculation of the bubble chain diagrams is performed, with the renormalon contribution providing an estimate of the magnitude of the power corrections.
To provide a numerical validation, we model the LCDAs of both the $\mathrm{D}$ meson and $\overline{\mathrm{B}}$ meson to check the size of the power corrections.
The results from the renormalon model suggest a correction of $22\%$ for the $\mathrm{D}$ meson and $7\%$ for the $\overline{\mathrm{B}}$ meson.
This indicates that it is necessary to account for power corrections when the quark is not sufficiently heavy in studies of the matching between QCD and HQET LCDAs.

\section*{Acknowledgement}
The authors would like to thank Qi-An Zhang, Jun Zeng, Zhifu Deng, and Yanbin Wei for sharing their calculation results with us. We also thank Yushan Su for valuable discussions and helpful comments on our results.
This work is supported by Natural Science Foundation of China under grant No.12125503 and 12335003. Jia-Lu Zhang is supported by T.D. Lee scholarship.

\appendix
\section{bubble chain calculation}
\label{app1}
In this section, we present the results for the one-loop bare bubble chain diagrams of both the QCD LCDA and the HQET LCDA. \textbf{HQET A} and \textbf{QCD A} correspond to diagram (a) in Fig.~\ref{fig:diagram}, while \textbf{HQET B} and \textbf{QCD B} correspond to diagram (b). The color factor $C_F$ and $\mu_0^{2(n+1)\epsilon}$ are omitted in the following results:
\begin{widetext}
\begin{itemize}
    \item \textbf{HQET A}
\begin{align}
m_H\varphi^+_a=-\frac{i g^2  4^{ (n+1)
   \epsilon } R_{\epsilon }^n \Gamma (n
   \epsilon +\epsilon ) (w-v)^{-2 (n+1)
   \epsilon -1}}{8\pi ^{2-\epsilon }\tilde{f}_H(\mu)n_+\cdot v \Gamma
   (n\epsilon +1)}\theta(\omega-\nu),
\end{align}
\item \textbf{QCD A}
\begin{align}
\phi_{a}=&i4^{(n+1) \epsilon } \frac{g^2 R_{\epsilon }^n m_Q^{-2(n+1)\epsilon}\Gamma (n \epsilon +\epsilon )}{8
   \pi ^{2- \epsilon }P_H^+ \Gamma (1+n \epsilon )} \frac{\Gamma (-2 (n+1) \epsilon ) \Gamma (n \epsilon
   +2)}{\Gamma (2-(n+2) \epsilon )} \delta(u-s)\notag\\
   &-i\frac{4^{ (n+1) \epsilon }g^2  
   m_Q^{-2 (n+1) \epsilon }
     \Gamma (n \epsilon +\epsilon
   ) }{8\pi ^{2-\epsilon}f_H\Gamma (n \epsilon
   +1)P_H^+} \frac{\bar{u}^{1+n\epsilon}}{\bar{s}^{1-(n+2) \epsilon }}\frac{\theta(u-s)}{(u-s)^{2 (n+1) \epsilon +1}},
\end{align}
\item \textbf{HQET B}
\begin{align}
 m_H\varphi^+_b=&-\frac{i g^2 \omega  (\epsilon
   -1) 4^{ (n+1) \epsilon } R_{\epsilon
   }^n \csc (\pi  (n+1) \epsilon )
   \left(\frac{\nu }{\omega}\right)^{\epsilon }
   (\nu  (\omega-\nu ))^{-((n+1) \epsilon
   )-1}}{8 \pi ^{1-\epsilon }\tilde{f}_H(\mu)n_+\cdot v \Gamma (2-\epsilon )}\theta(\nu-\omega)\notag\\
   &-\frac{i g^2  4^{ (n+1)
   \epsilon} R_{\epsilon }^n \Gamma (n
   \epsilon +\epsilon +1) (\omega-\nu )^{-2 (n
   \epsilon +\epsilon +1)} }{8\pi ^{2-\epsilon } \tilde{f}_H(\mu) n_+\cdot v}\Bigg((\omega-\nu ) \,
   _2\tilde{F}_1\Big(1,(n+1) \epsilon ;n
   \epsilon +2;-\frac{\nu }{\omega-\nu}\Big)\notag\\
   &+\nu  (n \epsilon +1) \,
   _2\tilde{F}_1\Big(1,n \epsilon +\epsilon
   +1;n \epsilon +3;-\frac{\nu }{\omega-\nu
   }\Big)\Bigg)\theta(\omega-\nu),
\end{align}
\item \textbf{QCD B}
\begin{align}
    \phi_b=&i \frac{g^2  4^{ (n+1)
   \epsilon }(m_Q^2)^{-(n+1)
   \epsilon }R_{\epsilon }^n}{8\pi ^{2-\epsilon}f_HP_H^+}  (1-s)^{(n+2)
   \epsilon } (1-u)^{n \epsilon +1} \Gamma
   (n \epsilon +\epsilon )
   \Bigg(\frac{(\epsilon -1)^2 \,
_2\tilde{F}_1\left(1,n \epsilon +\epsilon
   ;n \epsilon +2;\frac{s-s u}{s-u}\right)
   (u-s)^{-2(n+1)
   \epsilon }}{s-1}\notag\\
   &-(n+1) u \epsilon 
   (u-s)^{-2 (n \epsilon +\epsilon +1)} \,
   _2\tilde{F}_1\left(1,n \epsilon +\epsilon
   +1;n \epsilon +2;\frac{s-s
   u}{s-u}\right)\Bigg)\theta(u-s)\notag\\
   &+\frac{i g^2   4^{ (n+1)
   \epsilon} (m_Q^2)^{-(n+1)
   \epsilon } R_{\epsilon }^n}{8\pi^{1-\epsilon}f_H P_H^+ \Gamma
   (1-\epsilon )} u(1-s)^{(n+2)
   \epsilon } s^{-n \epsilon -1} \csc (\pi 
   (n+1) \epsilon ) (u-s u)^{-\epsilon } \notag\\
   &(s
   (-\epsilon )+s+u (\epsilon -1)+1)
   (u-s)^{-(n+1) \epsilon -1}\theta(s-u).
\end{align}
\end{itemize}

\end{widetext}

\end{document}